\documentclass[fleqn,usenatbib]{mnras}

\usepackage[T1]{fontenc}

\DeclareRobustCommand{\VAN}[3]{#2}
\let\VANthebibliography\thebibliography
\def\thebibliography{\DeclareRobustCommand{\VAN}[3]{##3}\VANthebibliography}

\usepackage{graphicx}	
\usepackage{amsmath}	
\usepackage{amssymb}	
\usepackage[normalem]{ulem}

\usepackage[dvipsnames]{xcolor}

\newcommand{\be}{\begin{equation}} 
\newcommand{\ee}{\end{equation}} 

\title[Are interactions shaping the jet?]{Are Interactions with Neutron Star Merger Winds Shaping the Jets?}

\author[L. Nativi et al.]{
L. Nativi$^{1}$\thanks{E-mail: lorenzo.nativi@astro.su.se},
G.~P. Lamb$^{2}$, 
S. Rosswog$^{1}$, C. Lundman$^{1}$, G. Kowal$^{3}$,
\\
$^{1}$Department of Astronomy and Oskar Klein Centre, Stockholm University, AlbaNova 10691 Stockholm, Sweden\\
$^{2}$School of Physics and Astronomy, University of Leicester, University Road, Leicester LE1 7RH, UK\\
$^{3}$Escola de Artes, Ci\^{e}ncias e Humanidades, Universidade de S\~{a}o Paulo, Av. Arlindo B\'{e}ttio, 1000 -- Vila Guaraciaba, \\CEP: 03828-000, S\~{a}o Paulo -- SP, Brazil
}

\date{Accepted XXX. Received YYY; in original form ZZZ}

\pubyear{2021}

\begin{document}
\label{firstpage}
\pagerange{\pageref{firstpage}--\pageref{lastpage}}
\maketitle

\begin{abstract}
Jets can become collimated as they propagate through dense environments and understanding such
interactions is crucial for linking physical models of the environments to observations. In this 
work, we use 3D special-relativistic simulations to study how jets propagate through the environment
created around a neutron star merger remnant by neutrino-driven winds. We simulate four jets 
with two different initial structures, top-hat and Gaussian, and two luminosities. After jet
breakout, we study the angular jet structures and the resulting afterglow light curves. 
We find that the initial angular structures are efficiently washed out during the propagation,
despite the small wind mass of only $\sim 10^{-3}$ M$_\odot$. 
The final structure depends on the jet luminosity as less energetic jets are more strongly collimated, and entrainment of baryons leads to a moderate outflow Lorentz factor ($\approx 40$).
{ Although our jets are not specifically intended to model the outflows of the GW170817 event, we show that they can be used to produce light curves consistent with} the afterglow observed in the aftermath of GW170817.
{ Using this procedure we show how the} inferred physical 
parameters e.g., inclination angle, ambient particle number density, can vary substantially 
between { independent fits of the same dataset}
and appear to be sensitive to smaller details of the angular jet shape, 
indicating that observationally inferred parameters may depend sensitively on the employed
jet models.
\end{abstract}

\begin{keywords}
gamma-ray bursts -- method: numerical -- hydrodynamics -- jets and outflows -- neutron star mergers -- relativistic processes
\end{keywords}


\section{Introduction}
\label{sec:intro}
Binary neutron star (BNS) mergers have long been suspected to
produce the central engines of short gamma ray bursts (sGRB) \citep{Eichler1989}. The link was firmly established in August 2017, after the combined 
detection of gravitational waves and a short GRB from the same BNS merger \citep{Abbott2017a, Abbott2017b, Abbott2017c, Goldstein2017, Savchenko2017}. The actual origin of the gamma-ray signal is still debated, coming from either an off-axis jet or from the shock breakout of a relativistic cocoon inflated by the jet itself \citep{Lundman2021, Gottlieb2018a, Gottlieb2018b, Kasliwal2017, Mooley2018a, Nakar2017}. Nevertheless, the multi-band observations of a rising afterglow \citep{Margutti2017, Margutti2018, Davanzo2018, Lyman2018, Lamb2019a, Alexander2017, Hallinan2017, Troja2017, Troja2018, Troja2019} together with the detection of superluminal motion \citep{Mooley2018b, Ghirlanda2019, Hotokezaka2019} have settled the presence of a jet that successfully broke out from the surrounding ejecta, observed off-axis with a viewing angle $\theta_{\rm obs}\approx 19^{\circ}$ \citep{Mooley2018a, Lazzati2018, Murguia-Berthier2017, Lamb2018b, Margutti2020}.
The information obtainable from afterglow observations is strongly dependent on the angular structure of the emerging jet. This structure might be mainly determined by the launching process \citep{Kathir2019a} or it may arise as a consequence of the interaction with the surrounding environment during propagation. Therefore, an understanding of the processes that shape the jet could in principle provide insights into both jet formation, and the post-merger environment. { Recent relativistic (magneto-) hydrodynamics simulations   \citep{Lazzati2021, Urrutia2021, Pavan2021, Geng2019, Gottlieb2020, Gottlieb2021a, Gottlieb2021b, Gottlieb2021c, Nathanail2021, Murguia-berthier2014, Murguia-Berthier2017, Murguia-Berthier2021a, Beniamini2020b, Hamidani2021} have illustrated the importance of the ambient medium in shaping the jet, thereby highlightning the importance of understanding the remnant structure and  ejecta properties.}
The joint events GW170817 and GRB170817A were followed by an additional electromagnetic transient spanning the spectral bands from UV to optical and IR on time scales from days to weeks \citep{Abbott2017b, Arcavi2017, Cowperthwaite2017, Evans2017, Drout2017, Pian2017, Smartt2017, SoaresSantos2017, Tanvir2017, Utsumi2017}. The observed properties were consistent
with the expectations for  a thermal transient powered by the radioactive decay of freshly synthesized r-process elements \citep[a so-called "macronova" or "kilonova" e.g.,][]{Li98, Kulkarni2005, Rosswog2005, Metzger2010,Roberts11, Kasen13a, Kasen2015, Kasen2017, Yu2013, Metzger2017, Tanaka2017, Perego2017, Rosswog2018}. Understanding the properties of this signal requires in-depth investigation of all the processes that can unbind material during and after a BNS merger, together with the available amount of free neutrons provided by each ejection channel. By now, several mass-ejection channels have been identified and they differ in terms of launch time, mass, electron fraction $Y_e$ and velocity. During the merger $\sim 10^{-3}-10^{-2}$ M$_{\odot}$ of material are ejected
dynamically \citep{Rosswog1998b, Rosswog1999, Rosswog2002a, Oechslin2007a, Bauswein13, Radice2018c}. On longer time scales { ($\sim 1$ s)} a few $10^{-2}$ M$_{\odot}$ of material can be unbound from the torus surrounding the remnant by the action of nuclear heating \citep{Fernandez2013, Fernandez2015, Just15, Metzger2008a}, magnetic \citep{Siegel2015, Siegel2017, Siegel2018, Ciolfi2017, Ciolfi2020, Murguia-Berthier2021b} and viscous effects \citep{Shibata2017, Shibata2019a, Radice2018b, Fujibayashi2018, Fujibayashi2020}. 
Weak interactions play a key role since they can change the
initially extremely low electron fraction and they are therefore of paramount importance for nucleosynthesis and the electromagnetic appearance of a BNS merger \citep{Ruffert1997a, Rosswog2002b, Dessart2009, Perego2014, Perego2017, Martin2015, Martin2018, Miller2019, Murguia-Berthier2021b}. 

In \citealt{Nativi2021} we investigated { how the propagation of a jet through a neutrino-driven wind \citep{Perego2014}
impacts the resulting radioactive transient. 
We found in particular that an outer layer of low-$Y_e$ material with only very small mass can have a significant impact on
the resulting electromagnetic signal, once more
underlining how crucial the realistic modelling of 
the jet environment is for interpreting observations \citep[see also][]{Lazzati2021, Pavan2021, Ito2021}. }
For example, a relativistic jet can punch away such low-$Y_e$, high-opacity material causing the macronova to { become} brighter earlier for observers close to the polar axis. \\
Here, we investigate the same neutrino-driven wind environment as in {\citealt{Nativi2021}}, { based on the simulations of \cite{Perego2014},} but from a different perspective. We want to know
to which extent the initial jet structure is preserved when a jet propagates through a realistic
post-merger environment. To this end
we perform high-resolution 3D special relativistic hydrodynamic simulations of jets with different initial angular structures and luminosities. In Section~\ref{sec:methods} we briefly discuss the numerical methods adopted for this study, and the simulation setup. We present our findings in Section~\ref{sec:results} { and, as a conclusive exercise, we attempt a comparison between the light curves produced by our resulting jet structures and the only afterglow with a certain connection to neutron star mergers i.e., GRB\,170817A.}
We summarize and list our main findings in Section~\ref{sec:discussion}.

\section{Methods}
\label{sec:methods}
We use the same special-relativistic hydrodynamic high-order shock-capturing adaptive mesh 
refinement Godunov-type code {\tt AMUN}\footnote{AMUN code is open source and freely available from
\url{https://gitlab.com/gkowal/amun-code}.}, as  in {\citealt{Nativi2021}}. 
In the present simulations the Riemann states at each cell interface are reconstructed using the 5$^{th}$-order Monotonicity Preserving (MP5) scheme of \cite{Suresh1997}, and the Harten-Lax-van Leer (HLL) Riemann solver \citep{Harten1983b} is used. The time integration is performed using the 3$^{\rm rd}$ order Strong Stability Preserving Runge-Kutta (SSPRK) algorithm from \cite{Gottlieb2011} with a value of 0.5 for the
Courant-Friedrichs-Lewy  number \citep{Courant_etal:1928}. The numerical integration of the special relativistic hydrodynamic equations requires nonlinear conversion of the conservative to primitive variables. In this work we used the $1D_W$ scheme with a Newton-Raphson iterative solver \cite[see, e.g.,][]{Noble2006} with the tolerance of $10^{-8}$ as the primitive variable solver.

\subsection{Simulations setup}
\label{sec:simulations}

We explore here how a jet is impacted by the interaction with the ambient
medium produced by a neutrino-driven wind. 
We use the  wind profiles of \cite{Perego2014} in all simulations. 
We embed the wind in a steeply decreasing power-law atmosphere that settles to a constant density plateau with $\rho_{\rm atm}=10^{-12}$ g cm$^3$. 
The background density has been chosen low enough to ensure that the final bow-shock produced by the jet 
carries a negligible amount of mass-energy. 
This requirement has been later verified at the end of each simulation.
\par The computational grid is shaped as a rectangular Cartesian box above the equatorial plane, with extensions $-4\times10^4 \leq ( x\rm{, }y ) \leq 4\times10^4$ km in the polar direction and $0 \leq z \leq 1.6 \times 10^5$ km in the axial direction. We used the properties of our mesh to set up a grid characterized by a superposition of two patches. 
One patch is kept fixed from the beginning and is used to cover the early jet propagation inside the wind, keeping the maximum available resolution ($2.4\times2.4\times3.2$ km). 
This patch has a radial extension of $R_{\rm p1}= 100$ km and an axial extension of $z_{\rm p1}=2000$ km. 
The second patch has an initial resolution { of four times that} of the first patch and a radial extension\footnote{{ We use the adaptive mesh refinement algorithm included in {\tt AMUN} to ensure that the difference in resolution between neighbouring computational blocks never exceeds a factor of 2.}} $R_{\rm p2}= 2000$ km. 
In the axial direction the second patch is used to dynamically follow  the jet head defining $z_{\rm p2} = (t - t_{0})c$ for all $z_{\rm p2}>z_{\rm p1}$, with $c$ the speed of light and $t_{0}$ the actual time at which the jet is launched.

\subsubsection{Jets}
\label{sec:jets}
As in {\citealt{Nativi2021}}, we inject the jet as a boundary condition from the lower boundary of the domain \citep{Gottlieb2021a, Gottlieb2020, Gottlieb2018a, Harrison2018, Geng2019, Mizuta2013, Mizuta2009}. 
{ Around the origin we move the lower boundary above the equatorial plane to a height of $z_{\rm 0}=40$ km, where injection occurs. In this way we remove the presence of the remnant, whose evolution is not of interest in the current work, but we keep within the domain the material around the rest of the equatorial plane. }
To inject the jet into the computational domain we specify five  functions: density $\rho_{\rm j}(\theta)$ and pressure $p_{\rm j}(\theta)$ in the local rest-frame and the three components of the velocity $\mathbf{v}_{\rm j}(\theta) = (v_{\rm j}^{\rm x},v_{\rm j}^{\rm y},v_{\rm j}^{\rm z})$ in the laboratory frame, $\theta$ is
the angle from the jet axis.
The geometry of the jet-injection region is characterized by a spherical radius from the source $r_{\rm 0}$ and an opening angle $\theta_{\rm j}$, with corresponding solid angle is $\Omega_{\rm j} = 2\pi(1-\cos\theta_{\rm j})$.
The intrinsic properties of the plasma flowing into the domain can be obtained from its luminosity $L_{\rm j}$, its velocity $v_{\rm 0}^{\rm r}$ with $r$ expressing the radial direction (or equivalently the Lorentz factor $\Gamma_{\rm 0}^{\rm r}$) and a description of the contribution of the internal energy $h$.
\noindent In this paper we present the results of four simulations, characterized by two luminosities ($L_{\rm j}=$10$^{50}$ and 10$^{51}$ erg s$^{-1}$) and two different initial structures: Gaussian (\texttt{gs51} and \texttt{gs50}) and top-hat (\texttt{th51} and \texttt{th50}). To set the geometry, we chose a height for the jet injection $z_{\rm 0}$ and an initial opening angle $\theta_{\rm j}=15^{\circ}$, consistent with the results from GRMHD simulations \citep{Kathir2019a}. In all the simulations the engine is active for $\Delta t_{\rm inj}= 100$ ms with constant luminosity, and decays exponentially after being shut-off on a time scale of 10 ms. The injected plasma at the top of the injection region is defined by an initial radial Lorentz factor $\Gamma_{\rm 0}=5$ and a specific enthalpy in the jet core $h_{\rm c}=30$, resulting in an asymptotic Lorentz factor $\Gamma_{\infty}=150$. All the simulations adopt a polytropic equation of state with adiabatic exponent 4/3, appropriate for relativistic gases.
\par For the initial jet structure  we start from the assumption that the luminosity per unit solid angle
\be
\frac{dL}{d\Omega} = r^2 v_{\rm 0} \Gamma_{\rm 0}^2 \rho c^2 h,
\label{eq:dLdOmega}
\ee
\noindent is the same for an observer along the axis for any structure. 
We call $r = z_0/\cos\theta$ the injection radius i.e., the distance from the origin.
We proceed with a few simplifying assumptions: the local speed is radial, $v_0 \neq v_0(\theta)$, $\rho_{\rm j} \neq \rho_{\rm j}(\theta)$ (for fixed $r$) while
the enthalpy 
\begin{equation}\label{eq:h}
    h = 1 + \frac{4p_{\rm j}}{\rho_{\rm j}},
\end{equation}
\noindent is allowed to have an angular dependence.
For a top-hat jet, $h = h_{\rm c}$ is constant and for a Gaussian jet, $h(\theta = 0) = h_{\rm c}$ i.e., the Gaussian jet has the same $h$ along its axis as the top-hat jet. 
In this approach we are assuming that the whole inner structure can be ascribed to the thermal energy of the jet.
The total jet luminosity is then
\be
L_{\rm j} = \int \frac{dL}{d\Omega} d\Omega = 2\pi \int \frac{dL}{d\Omega} d\mu,
\ee
\noindent where $\mu \equiv \cos\theta$.

For a top-hat jet the luminosity per unit solid angle is constant for $\theta < \theta_j$, and the total luminosity is
\be
L = 2\pi \int \frac{dL}{d\Omega} d\mu = 2\pi (1 - \mu_j) \frac{dL}{d\Omega},
\ee
\noindent where $\mu_j = \cos\theta_j$. 
We can then solve for $\rho_{\rm j}$ at a given $r$ using Equation~\ref{eq:dLdOmega},
\be
\rho_{\rm j} = \frac{L_{\rm j}}{2\pi(1-\mu_j) r^2 v_0 \Gamma_0^2 h_c c^2},
\label{eq:rho}
\ee
\noindent and find the pressure, $p_{\rm j}$, from Equation~\ref{eq:h},
\be
p_{\rm j} = \frac{1}{4} (h_{\rm c} - 1) \rho_{\rm j} c^2.
\label{eq:p}
\ee
\noindent The primitive variables $(v, \rho, p)$ are thus fully specified for the top-hat jet.

For the Gaussian case we start again with our assumptions $v_0 \neq v_0(\theta)$ and $\rho_{\rm j} \neq \rho_{\rm j}(\theta)$. 
Therefore, using the same equation for $\rho$ (Equation~\ref{eq:rho}), we only need to find the new equation for $p = p(\theta)$ that corresponds to a Gaussian jet with the same pressure along the axis as the top-hat jet, and also carries the same total luminosity.

\noindent We want $h$ to have a Gaussian profile, therefore we assume
\be
h(\theta) - 1 = (h_{\rm c} - 1) \exp\left(-\frac{\theta^2}{\theta_j^2}\right).
\label{eq:h_gauss}
\ee

So, to find the primitive variables for a Gaussian jet, $v = v_0$, $\rho$ is computed from Equation~\ref{eq:rho} (same as for the top-hat jet) and $p$ is found by taking $h$ from Equation~\ref{eq:h_gauss} and subbing into Equation~\ref{eq:p}. 
The top-hat and Gaussian jet shapes are shown in Figure~\ref{fig:angprofiles}.

\section{Results}
\label{sec:results}
\begin{figure*}
    \centering
	\includegraphics[width=0.80\textwidth]{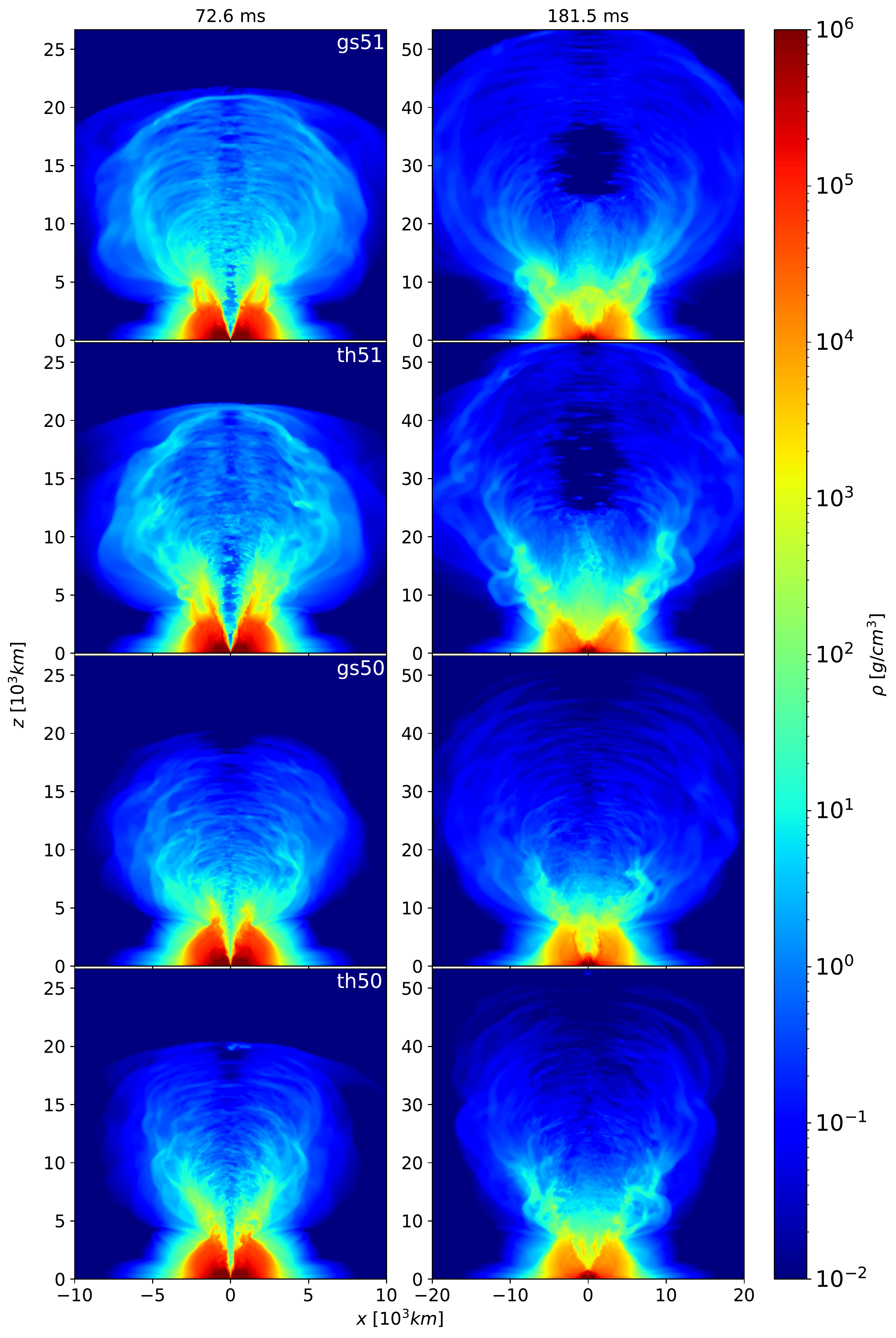}
	\caption{Vertical slices ($y=0$) of the rest-mass density shortly after the breakout (left) and after the jet shut-off (right) for all models. In all cases the jet punches a hole into the surrounding wind roughly proportional to its luminosity, consistently with what shown in \citealt{Nativi2021}. In the lateral direction the wind keeps expanding homologously. Particularly in the two high luminosity models (upper figures) some low density material is preceded by some at higher density. The material sitting on the top is injected before the breakout and has experienced more mixing. The matter injected after the breakout is only marginally affected by the interaction and maintains a lower density. These effects appear being more important in the Gaussian model, and are also visible in Fig.~\ref{fig:multilfac} and Fig~\ref{fig:multienth}.}
    \label{fig:multidens}
\end{figure*}

\begin{figure*}
    \centering
	\includegraphics[width=0.8\textwidth]{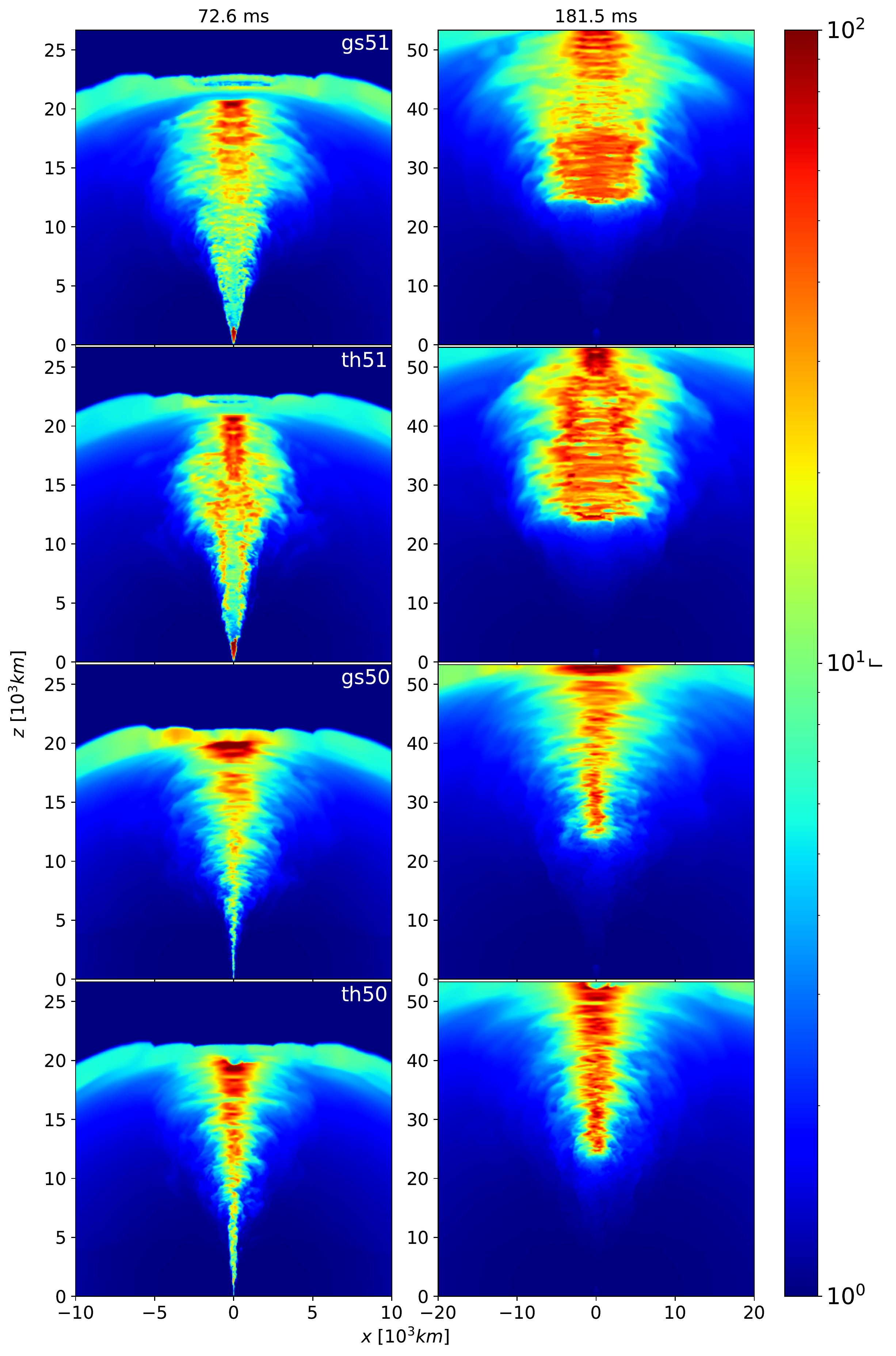}
	\caption{Same as in Fig~\ref{fig:multidens} but for the Lorentz factor. All jets experience a first strong collimation shock converting the specific kinetic energy into thermal and then several recollimation events (see Fig~\ref{fig:multienth}). After the breakout the jet drives a strong relativistic shock in the surrounding environment. At later times the jet material keeps being adiabatically accelerated towards its asymptotic value $\Gamma h$. The material injected before the breakout is affected by baryon pollution and is characterised by an averagely lower value for $\Gamma$, as described in Fig~\ref{fig:multidens}.}
    \label{fig:multilfac}
\end{figure*}

\begin{figure*}
    \centering
	\includegraphics[width=0.8\textwidth]{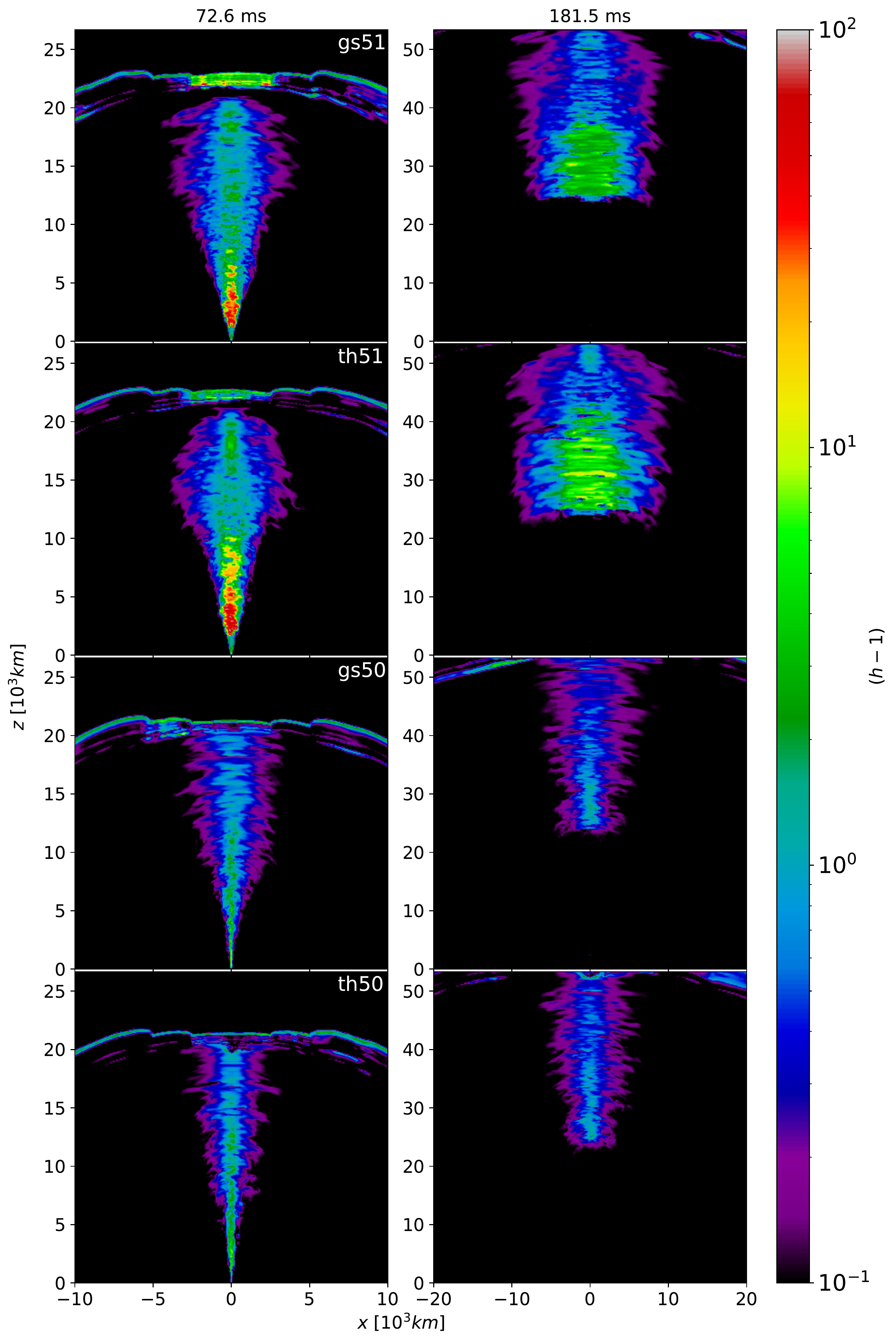}
	\caption{Same as in Fig.~\ref{fig:multidens} and  Fig.~\ref{fig:multilfac} but showing the specific enthalpy $h-1$. These plots together with those of the Lorentz factor describe the evolution of the jet. It is clear how the most polluted material is characterised by both a lower $\Gamma$ and a lower $h-1$, meaning a sensitive reduction of $\Gamma h$ due to mass entrainment for the material injected before the breakout.}
    \label{fig:multienth}
\end{figure*}

Overall, we find the dynamical evolution of the system consistent with the expectations from { other results} \citep{Mizuta2009, Bromberg2011, Mizuta2013, Nagakura2014, Murguia-berthier2014, Murguia-Berthier2021a, Harrison2018, Gottlieb2018a, Gottlieb2020, Gottlieb2021a, Gottlieb2021b, Duffell2018, Nativi2021, Urrutia2021, Ito2021, Hamidani2020, Hamidani2021}. 
In all models, the jets are quickly engulfed by a cocoon and experience several recollimation shocks before they manage to break out from the ejecta. 
Figures~\ref{fig:multidens}, \ref{fig:multilfac} and \ref{fig:multienth} show the evolution of all models for rest-mass density, Lorentz factor and $(h-1)$ respectively.
After breaking out, all the 
{ previously collimated jets start expanding sideways widening a funnel with an opening angle} close to the initial $\theta_j$. 
Such angular extension is also visible in the final angular profiles and is consistent with the expectations \citep[e.g.][]{Mizuta2013, Nagakura2014, Harrison2018}, see Fig.~\ref{fig:angprofiles}.
The post-breakout evolution proceeds with a progressive sideways expansion of the jet accompanied { by} a gradual decreasing of $(h-1)$ and an increasing Lorenz factor $\Gamma$.
We run all simulations until the jet reaches a configuration of ballistic expansion, with $(h-1)< 1$ everywhere { on} the grid. 
After this point, essentially all the internal energy in the plasma has been converted to kinetic energy, and the system undergoes free expansion. 
For all models we choose to extract the angular profiles at $\approx0.4$ s.
\par

{In order to use the final jet structure to produce afterglow light curves, we need to estimate the properties of the forward afterglow shock. Specifically, we need the energy per solid angle d$E(\theta)$/d$\Omega$, and the Lorentz factor of the shock as a function of angle $\overline{\Gamma}(\theta)$. At late times, the whole jet has been compressed into a thin layer behind the forward shock. Therefore, we estimate these properties by radially averaging the relevant quantities, while keeping their angular dependences. We use mass-averages in order to conserve both mass and energy in the averaging process.}

{However, the averaging process should not consider all plasma inside the grid. 
For instance, slow, heavy plasma that flows behind the jet will not have time to catch up and contribute to the shock on relevant time scales. 
We therefore need a procedure to estimate which parts of the jet are likely to contribute to the afterglow shock at relevant times. 
Here, we simply adopt a threshold value of $\Gamma h = 2$; fluid with a lower value is assumed to not contribute much to the afterglow\footnote{{Afterglow lightcurves calculated using a threshold $\Gamma h = 5$, for the Lorentz factor profile, show no significant deviation when compared to those found using our adopted limit for fits to GW170817 data, see \S \ref{sec:GW170817}. Low-$\Gamma$ and low-energy components, that constitute the wider cocoon, will contribute to the afterglow for a mildly inclined system at the afterglow onset time, and for observers at much higher system inclinations will become the dominant emission component.}}. This value ensures that the mass-averaged Lorentz factor is not affected much by the slow plasma that follows the jet, while retaining plasma at fairly large angles to the jet axis.} 
We perform a cylindrical average of the 3D grid returned by the hydrodynamic simulation: this reduces the problem to 2D in the $(r,z)$ plane and with uniform spacing $\Delta r$, $\Delta z$.
The physical quantities we need { are:
\begin{equation}
    D = \Gamma\rho\; \mbox{,} \quad
    h = 1 + \frac{4 P}{\rho c^2}\; \mbox{,}\quad
    E = D \Gamma h c^2 - p - D c^2
\end{equation}
\noindent at every grid point.}
We assume the polar angle $\theta$ being distributed with uniform spacing $\Delta \theta$ in the interval $[0,\pi/2]$, then iterate over the grid.
{ As motivated above,} we adopt a threshold { value of} $\Gamma h = 2$ to allow the selection of relativistic material from the jet and cocoon, which may contribute to the final afterglow emission.
For each selected cell we then compute the corresponding polar angle for the angular bin $[\theta_l ; \theta_{l+1}]$. 
Where for each angular bin we define two variables:
\begin{align}
    & \Delta E_l = \Delta E_l + E_{ij} \mathcal{V}_{ij}\\
    & \Delta M_l = \Delta M_l + D_{ij} \mathcal{V}_{ij} \\
    & \Delta \Gamma_l = \Delta \Gamma_l + \Gamma_{ij} D_{ij}  \mathcal{V}_{ij}
\end{align}
\noindent where $\mathcal{V}_{ij} = 2\pi r_i \Delta r \Delta z$ is the volume of cell $\{i,j\}$.
Finally, we can get the two required quantities as:
\begin{align}
    & \frac{\rm{d}E(\theta)}{\rm{d}\Omega} \approx \frac{\Delta E_l}{\Delta \Omega_l} = \frac{\Delta E_l}{2\pi \lvert \Delta\mu \rvert}\ \rm{, with }\quad \mu = \cos (\theta),\\
    & \overline{\Gamma}(\theta) \approx \frac{\Delta \Gamma_l}{\Delta M_l}.
\end{align}
The resulting profiles are shown in Fig~\ref{fig:angprofiles}.
Clearly, there are structural differences, but in all cases the initial jet structure is largely washed out in the interaction with the ejecta despite their relatively small amount of mass ($\sim 10^{-3}$ M$_{\odot}$). 

The Gaussian jets have a resultant profile with a broader low energy wing and they are less energetic on-axis, while the top-hat jets result in a profile that is slightly more energetic and faster at small angles suggesting that for such initial jet structures it is easier to preserve the energetic spine.
In all models we find that, due to baryon entrainment, the final average Lorentz factor
is systematically lower than their asymptotic value  ($\Gamma_{\infty} = \Gamma h \approx 150$).
In the late time maps, shown on the right of Fig.~\ref{fig:multienth}, the material just below the head is characterised by a much lower specific enthalpy than the material injected later.
All the matter that is flowing into the domain before the jet breakout is heavily affected by the interaction with its surroundings. 
This interaction triggers hydrodynamic instabilities and
turbulence at the jet-ejecta interface, favouring mixing and mass entrainment.
The material injected after the jet breakout is, however, only marginally affected by the instabilities at the interface, and this material proceeds along the opened up funnel retaining most of its energy.

\subsection{Application to the GRB\,170817A afterglow}\label{sec:GW170817}

\begin{figure*}
    \centering
	\includegraphics[width=0.85\textwidth]{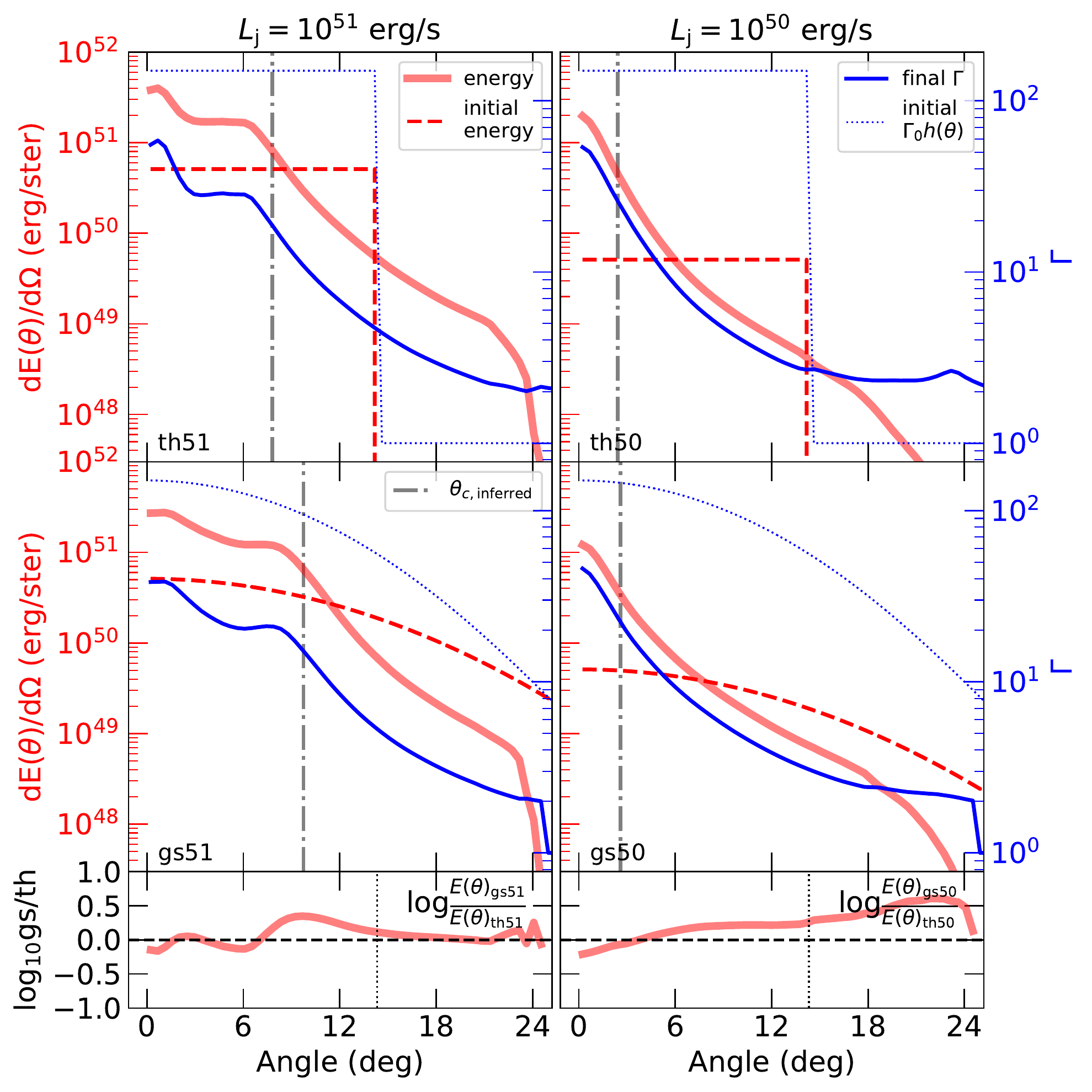}
	\caption{The resultant energy (solid red line) and Lorentz factor (solid blue line) profiles with polar angle for each model at time t $\approx 0.4$s.
	The top four panels include the injected energy (dashed red) and Lorentz factor (dotted blue) profiles for each model.
	The bottom panels show the ratio of the resultant energies with angle between an initially Gaussian profile (middle panels) and an initially `top-hat' profile (top panels) jet.
	For each profile, $\theta_j = 0.25 \sim 15^\circ$, marked in the bottom panel by a vertical dotted line, the horizontal dashed line indicates \texttt{gs}/\texttt{th}$=1$.
	The resultant energy and Lorentz factor profiles are calculated by summing cells with $\Gamma h\geq 2$. 
	The jet opening angle as inferred from the jet-break timing for on-axis afterglow light curves is shown for each model as a vertical dash-dotted grey line.
	}
    \label{fig:angprofiles}
\end{figure*}

The resulting energy and Lorentz factor jet profiles will be effectively frozen after break out and the jet becomes conical, expanding as a function of distance from the central engine.
This resultant jet structure can be used { to inform the choice of} fiducial energy and Lorentz factor profiles for structured jet afterglow models, such as those used to model the late time afterglow to GRB\,170817A \citep[e.g.][]{Lyman2018, Troja2018, Troja2019, Lamb2018a, Lamb2019a, Resmi2018, Ryan2020, Salafia2019, Fernandez2021}.

\begin{figure*}
\centering
    \includegraphics[width=\textwidth]{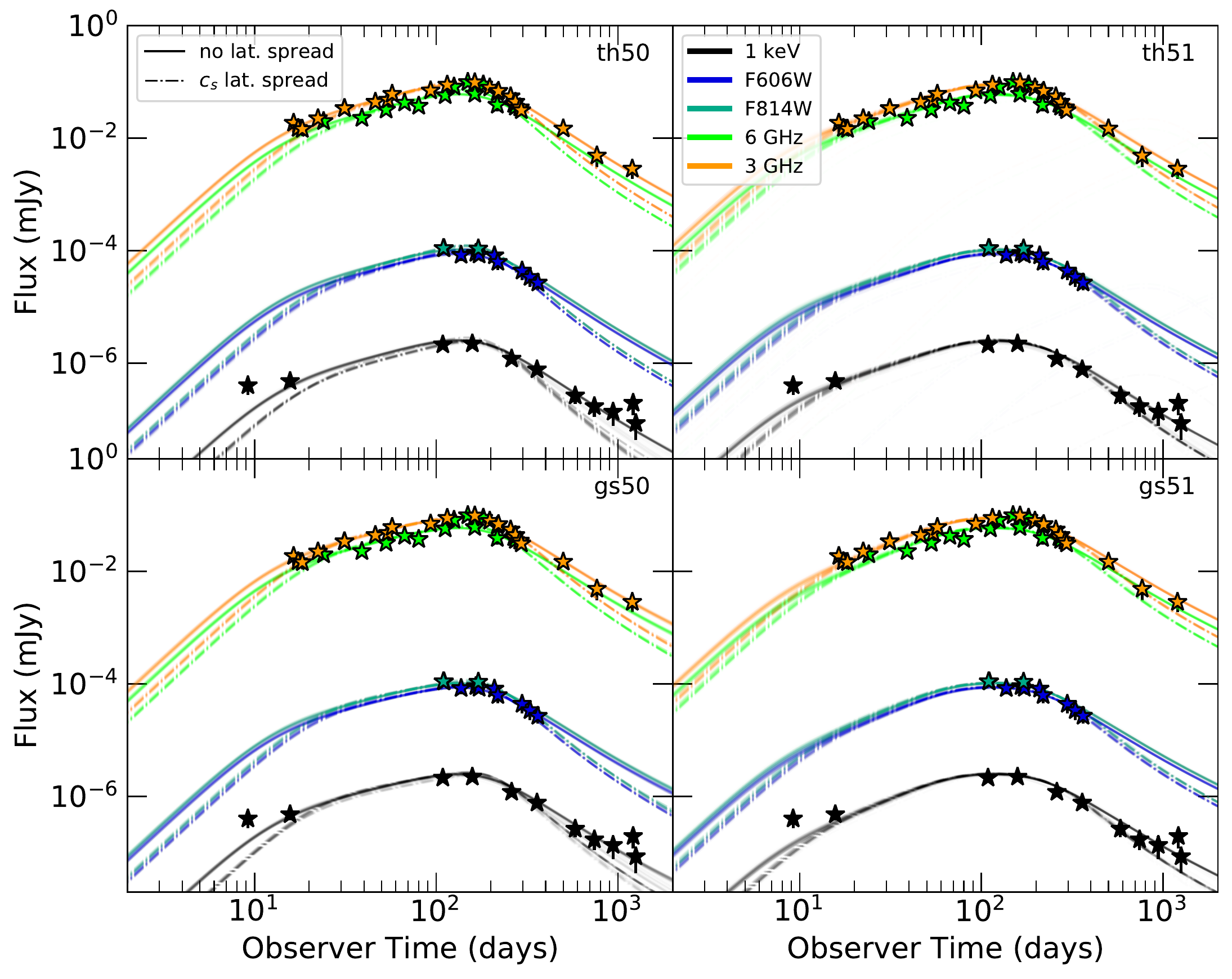}
    \caption{100 randomly selected afterglow lightcurves from the posterior distributions of MCMC fits to GRB\,170817A data for each jet profile where 50 lightcurves include the effects of sound speed lateral spreading (dash-dotted), and 50 are shown without lateral spreading (solid). The energy and Lorentz factor profiles are fixed to those from our simulations and the inclination, ambient density and microphysical parameters are allowed to vary.}
    \label{fig:GRB170817A}
\end{figure*}

We use the energy and Lorentz factor profiles  shown in Fig. \ref{fig:angprofiles} with the afterglow models developed in \cite{Lamb2017, Lamb2018b, Lamb2021} to predict and model the appearance of GRB afterglow lightcurves from structured jets at any inclination.
{ Although it is not our intention to infer the afterglow properties of GRB\,170817A, 
{we use this event's afterglow light curve to demonstrate
that our resultant jet profiles can nevertheless 
 produce afterglows consistent with the expectation from structured jet models}
.}
We use two cases for each model, without lateral spreading, where the jet is assumed to have no sideways expansion i.e., perpendicular to the radial direction, and with lateral spreading, where we assume sideways expansion at the local sound-speed of the jet element \citep[see][for details]{Lamb2018b,Lamb2021}.
We use the X-ray, optical, and radio frequency data-sets as presented in \cite{Troja2021, Balasubramanian2021, Makhathini2020} and references therein\footnote{Note that in all cases, the very late time X-ray and 3\,GHz radio data, as presented here, can be accommodated by the GRB afterglow model with either no, or moderate, lateral spreading -- where sound-speed lateral spreading is assumed to represent an upper limit on the sideways expansion.}.
As the energy and initial Lorentz factors are set by our resultant profiles, we fix the electron index, $p=2.15$, following standard closure relations for the afterglow spectrum, and adopt a redshift, $z=0.009783$ \citep{Levan2017} with Planck cosmology giving a distance, $D_L\sim43.7$\,Mpc.
We fit for the inclination, $\iota$, the ambient particle number density, $n$,
and the microphysical parameters, $\varepsilon_B$ and $\varepsilon_e$.
The best fit parameters and lightcurves are found via a Markov Chain Monte Carlo (MCMC) using \texttt{emcee} \citep{mackey2013}, with flat log priors for each parameter except the inclination, which has a flat prior in $\cos\iota$.
The prior ranges and the central plus 16th and 84th percentile values are shown in Table \ref{tab:GW170817}.
A sample of lightcurves drawn randomly from the posterior distribution are shown in Fig. \ref{fig:GRB170817A} with the observed data points, where the models without lateral spreading have solid lines, and those with maximal lateral spreading, have dash-dotted lines.

\begin{table*}
    \centering
    \caption{Free parameters for the MCMC model lightcurve fits to GW170817/GRB\,170817A afterglow data. All priors are flat in the range shown. Each model profile is fit with no lateral spreading, and with sound-speed lateral spreading. { Note that the model jet structures are found via simulations which are inconsistent with the conditions in GW170817/GRB\,170817A, and as such, the fit afterglow parameter values should not be used to directly model or infer the properties of this event.}}
    {\begin{tabular}{cccccc}
          Model & Lat. spread (y/n) & $\iota$ (rad) & $\log\varepsilon_B$ & $\log\varepsilon_e$ & $\log n$ $\log$(cm$^{-3}$) \\
          \hline
          \hline
          Prior & -- & $[\cos(0.75),~\cos(0.25)]$ & $[-6,~-0.3]$ & $[-6,~-0.3]$ & $[-6,~2]$ \\
          th50 & n &$ 0.29 ^{+ 0.04 }_{- 0.03 }$  & 
$ -2.31 ^{+ 0.62 }_{- 0.74 }$  & 
$ -1.54 ^{+ 0.52 }_{- 0.42 }$  & 
$ -3.12 ^{+ 0.49 }_{- 0.34 }$  \\
          th50 & y &$ 0.25 ^{+ 0.01 }_{- 0.00 }$  & 
$ -0.59 ^{+ 0.05 }_{- 0.07 }$  & 
$ -2.20 ^{+ 0.04 }_{- 0.03 }$  & 
$ -4.03 ^{+ 0.06 }_{- 0.03 }$\\
          gs50 & n &$ 0.41 ^{+ 0.05 }_{- 0.05 }$  & 
$ -3.24 ^{+ 0.82 }_{- 0.65 }$  & 
$ -0.93 ^{+ 0.49 }_{- 0.62 }$  & 
$ -2.17 ^{+ 0.35 }_{- 0.44 }$\\
          gs50 & y &$ 0.28 ^{+ 0.03 }_{- 0.02 }$  & 
$ -1.01 ^{+ 0.33 }_{- 1.09 }$  & 
$ -1.89 ^{+ 0.82 }_{- 0.16 }$  & 
$ -3.96 ^{+ 0.38 }_{- 0.21 }$ \\
          th51 & n &$ 0.45 ^{+ 0.02 }_{- 0.02 }$  & 
$ -4.37 ^{+ 1.14 }_{- 1.10 }$  & 
$ -1.60 ^{+ 0.75 }_{- 0.79 }$  & 
$ -1.36 ^{+ 0.20 }_{- 0.19 }$\\
          th51 & y &$ 0.37 ^{+ 0.02 }_{- 0.01 }$  & 
$ -2.78 ^{+ 0.96 }_{- 1.48 }$  & 
$ -1.85 ^{+ 1.06 }_{- 0.86 }$  & 
$ -2.66 ^{+ 0.15 }_{- 0.17 }$\\
          gs51 & n &$ 0.51 ^{+ 0.03 }_{- 0.02 }$  & 
$ -4.29 ^{+ 1.09 }_{- 1.15 }$  & 
$ -1.65 ^{+ 0.70 }_{- 0.76 }$  & 
$ -1.01 ^{+ 0.18 }_{- 0.18 }$\\
          gs51 & y &$ 0.43 ^{+ 0.01 }_{- 0.01 }$  & 
$ -3.47 ^{+ 1.03 }_{- 1.23 }$  & 
$ -1.73 ^{+ 0.83 }_{- 0.74 }$  & 
$ -2.15 ^{+ 0.13 }_{- 0.14 }$ \\
\hline
    \end{tabular}
    \label{tab:GW170817}}
\end{table*}

For each of our models we use an afterglow lightcurve without lateral spreading, and with $\iota=0$ to measure the jet break time as seen by a distant observer.
By fitting a simple smoothly broken power law function for the model afterglow light curve, as in \cite{Lamb2021}, we find the best-fit observed jet-break time in each case, we then use the known energy in the jet and the fixed ambient number density to estimate the effective core angle of each profile.
The inferred $\theta_c$ angles are shown on Fig. \ref{fig:angprofiles} as vertical grey dash-dotted lines in each panel.
The core values in each case are:
$\theta_c = [0.042,~0.045,~0.136,~0.170]$\,rad for \texttt{th50}, \texttt{gs50}, \texttt{th51}, \texttt{gs51} respectively.
These jet core angles as estimated from the `observed' jet break time imply that, for a fixed merger environment density into which a jet is launched and for a fixed initial jet profile, that the opening/core angle of the resultant jet depends on the initial jet power.
We note that the peak energy for the resultant profiles, in each case, is at a comparable level, $\sim$few $10^{52}$ erg, in terms of isotropic equivalent energy.

Using the inferred core angles for each profile, we find the ratio $\iota/\theta_c$ for the posterior distribution of inclination angles in each case.

As can be seen in Table \ref{tab:GW170817}, where lateral spreading is included, the preferred inclination of the system is reduced and as $\theta_c=$constant in each case, then spreading models have a typically smaller ratio $\iota/\theta_c$ than the non-spreading models.
{ Where superluminal motion is included in combination with afterglow modelling for GRB\,170817A, \cite{Nakar2021} found that the ratio of the inclination, $\iota$, to the jet core angle, $\theta_c$, should fall in the range $4\lesssim\iota/\theta_c\lesssim6$, to be consistent with all the observations.}
{ We did not include the radio imaging constraints in our MCMC and, as such,} none of our models fits comfortably in this range. 
{ However, for}
the low- and high-energy models at $L_{\rm j}=10^{50-51}$ erg s$^{-1}$, 
{ the jet structure has little impact on the ratio $\iota/\theta_c$, which depends more significantly on the initial power of the jet, and whether lateral spreading is included or not.
Where lateral spreading is included, the $\iota/\theta_c$ values are marginally smaller in all cases, consistent with the findings of \cite{Fernandez2021}.}
It is, however, not  the scope of this paper to consistently fit the afterglow to GRB\,170817A.
We additionally note that, despite the differences in the resultant profiles for both fixed injected energy and fixed initial jet structure, the afterglow lightcurves can all reproduce the observed temporal behaviour of the late afterglow to GW170817/GRB\,170817A.
However, we note that although all of these resultant jet profiles can produce afterglow lightcurves that fit the data, and all have nearly consistent ratios, $\iota/\theta_c$, for fixed jet energies, the value of the other fit parameters varies significantly between the models.
{ For GRB afterglow models, there are some well known parameter degeneracies and as we employ fixed energy/Lorentz factor profiles for each model, where all other physics is kept the same, the variation in preferred parameters between the different structure profiles is likely a result of these e.g., the deceleration timescale is proportional to the ratio $(E/n)^{1/3}$, and the inferred inclination angle can depend on the jet core size or structure profile, for more details see \cite{Beniamini2020a, Ryan2020, Nakar2021} and, where intrinsically more diverse jet structure profiles are considered, \cite{Takahashi2021}.}
For a jet with $L_{\rm j}=10^{50}$ erg s$^{-1}$, where the initial profile is a top-hat jet, the inferred inclination for the data is $\iota\sim0.25\pm0.01$\,rad ($14.3\pm0.6$ degrees) whereas for an initially Gaussian jet structure, $\iota\sim0.28\pm0.03$ ($16.0\pm1.8$ degrees) showing that, where we use the resultant angular profiles of realistic simulations following jet propagation through a neutron star merger environment, then the assumed injected jet structure can influence the inferred parameters from the subsequent afterglow fitting.
This has implications for Hubble parameter estimates from GW counterparts where simulations and/or afterglow model fits are used \citep[e.g.][]{Hotokezaka2019, Wang2021, Mastrogiovanni2021}.

\section{Conclusions}
\label{sec:discussion}
We have investigated to which extent a jet from a BNS merger maintains its initial structure when plowing through the matter previously ejected through a neutrino-driven wind, and the consequences for the observed afterglow emission. To achieve this, we have performed a set of 3D special-relativistic hydrodynamic simulations where two different profiles for the initial jet structure were adopted, and two different luminosities. We propagated the jets in a previously simulated neutrino-driven wind environment (which was also used in \cite{Nativi2021}). The final angular jet profiles were recovered after the jets broke out of the ejecta, and the profiles were used to compute afterglow light curves.
{ Our study was not designed to model GW170817, nevertheless, since this is so far the only observed multi-messenger event from a merging neutron star, it is tempting to ask if the systems we have simulated can produce an afterglow lightcurve that is consistent with that from GW170817/GRB\,170817A.
Interestingly, we find that all of our resultant jet structures can reproduce the observed, multi-band afterglow light curve of this event.
}\\
Our main findings are:
\begin{itemize}
    \item Despite the relatively small amount of material in such winds ($\sim 10^{-3}$ M$_{\odot}$) the emerging jet appears to be entirely shaped by the interaction. 
    Hence jets initially injected with a different structure possess a  similar shape at the end of the simulations.
    
    \item Mixing is important for the final Lorentz factor profiles. Baryon entrainment is a consequence of hydrodynamic instabilities and turbulence arising at the contact interface between the jet and ejecta, and cannot be properly represented in 2D simulations \citep{Duffell2018, Harrison2018, Urrutia2021}. We find that mixing is important for the part of the jet injected before the jet breakout, while the latter part of the jet is barely affected, retaining its initial composition and energy per baryon { \citep[see][for more details]{Harrison2018, Gottlieb2020, Gottlieb2021a, Gottlieb2021b}.}

    \item { We find that the Lorentz factor, $\Gamma$, for all our jet models peaks at a value consistent with or above the theoretical minimum for the short GRB population, $\Gamma\approx40$ \citep{Nakar2007}.}
    
    \item We further explore the jet core angles, $\theta_{\rm c}$, that would be inferred (via the jet break time) given an observation of the afterglow. These angles show a dependency on the injected jet power. This is a consequence of the different degree of collimation: low energy jets are more effectively collimated by the interaction with the environment than high energy jets.
    
    \item The final jet profiles differ in their small scale features, which are likely sensitive to simulation details such as the method of jet launching, numerical resolution  as well as the properties of the surrounding environment. Smaller features of the jet structure can therefore not be constrained by current jet simulations, but they significantly change our best fit values of the afterglow parameters. Where single simulation results are used to model the afterglow (i.e. to get precise inclination angles for use in cosmology as in e.g., \citealt{Hotokezaka2019, Wang2021}), the small differences arising from the choice of the initial conditions and subsequent evolution impact the parameter values inferred by the best fit.
    
\end{itemize}

\section*{Acknowledgements}
This work has been supported by the Swedish Research 
Council (VR) under grant number 2016- 03657\_3, 
by the Swedish National Space Board under grant number 
Dnr. 107/16, 
the research environment grant 
``Gravitational Radiation and Electromagnetic Astrophysical
Transients (GREAT)" funded 
by the Swedish Research 
council (VR) under Dnr 2016-06012 and 
by the
Knut and Alice Wallenberg Foundation under Dnr KAW 2019.0112. 
We gratefully 
acknowledge stimulating interactions from COST Action CA16104 
``Gravitational waves, black holes and fundamental physics" (GWverse) and from COST Action CA16214 ``The multi-messenger physics and astrophysics of neutron stars" (PHAROS).\\
We acknowledge support from the Swedish National Space Agency.\\
The simulations were performed on resources
provided by the Swedish National Infrastructure for
Computing (SNIC) at Beskow and Tetralith
and on the resources provided by the North-German
Supercomputing Alliance (HLRN).\\
G.P.L. is supported by the Science and Technology Facilities Council, UK via grant ST/S000453/1.\\
G.K. acknowledges support from the São Paulo Research Foundation, FAPESP (grants 2013/10559-5 and 2019/03301-8).\\
The authors thank the anonymous referee for their timely and constructive comments that have improved the paper.

\section*{Data Availability}
The data used to produce the observational findings, i.e. the angular profiles describing the jets structures, are available from the author, L.N., upon reasonable request.
 



\bibliographystyle{mnras}
\bibliography{lnativi} 








\bsp	
\label{lastpage}
\end{document}